\documentclass[12pt]{article}
\usepackage{graphicx}

\newcommand\pubnumber{SNSN-323-63}
\newcommand\pubdate{\today}

\def\institute{DESY\\
Deutsches Elektronen-Synchrotron, D-22607 Hamburg, GERMANY}

\def\Title#1{\begin{center} {\Large #1 } \end{center}}
\def\Author#1{\begin{center}{ \sc #1} \end{center}}
\def\Address#1{\begin{center}{ \it #1} \end{center}}

\newcommand\pubblock{\rightline{\begin{tabular}{l} \pubnumber\\
         \pubdate  \end{tabular}}}
\newenvironment{Abstract}{\begin{quotation}  }{\end{quotation}}
\newenvironment{Presented}{\begin{quotation} \begin{center} 
             PRESENTED AT\end{center}\bigskip 
      \begin{center}\begin{large}}{\end{large}\end{center} \end{quotation}}

\def\beq{\begin{equation}}
\def\eeq#1{\label{#1}\end{equation}}
\def\eeqn{\end{equation}}

\def\beqa{\begin{eqnarray}}
\def\eeqa#1{\label{#1}\end{eqnarray}}
\def\eeqan{\end{eqnarray}}

\let\bar=\overbar

\def\Dslash{\not{\hbox{\kern-4pt $D$}}}
\def\dslash{\not{\hbox{\kern-2pt $\del$}}}

\def\msb{{\bar{\ssstyle M \kern -1pt S}}}

\def\ttbar{$t\bar{t}$}
\def\ee{$e^+e^-$}
\def\emu{$e^\pm\mu^\mp$}
\def\mumu{$\mu^+\mu^-$}
\def\pp{$pp$}
\def\ttpt{$p_T^{t\bar{t}}$}
\def\ttm{$m_{t\bar{t}}$}
\def\tty{$y_{t\bar{t}}$}
\def\toppt{$p_T^{t}$}
\def\topy{$y_{t}$}
\def\met{\mbox{$\not \!\! E_T$}}

\begin{document}
\begin{titlepage}
\pubblock

\vfill
\Title{Measurement of the differential cross section\\
for top-quark-pair production in the dilepton channel\\
at $\sqrt{s}$ = 13 TeV with the CMS detector\\}
\vfill
\Author{Mykola Savitskyi on behalf of the CMS collaboration}
\Address{\institute}
\vfill
\begin{Abstract}
Measurements of normalized differential top-quark-pair ($t\bar{t}$) production cross sections are performed
using final states with two leptons ($e^+e^-$, $\mu^+\mu^-$, and $e^\pm\mu^\mp$) in proton-proton ($pp$) collisions at $\sqrt{s}$ = 13 TeV at the CERN LHC.
The data were recorded in 2015 with the CMS detector and correspond to an integrated luminosity of 2.2 fb$^{-1}$.
The $t\bar{t}$~production cross section is measured as a function of kinematic properties of the top quarks and the $t\bar{t}$~system in the full phase space, as well as of the jet multiplicity
in the event in the fiducial phase space. Several perturbative quantum chromodynamics (QCD) calculations are confronted with the data and are found to be broadly in agreement with the measured results.
\end{Abstract}
\vfill
\begin{Presented}
$9^{th}$ International Workshop on Top Quark Physics\\
Olomouc, Czech Republic,  September 19--23, 2016
\end{Presented}
\vfill
\end{titlepage}
\def\thefootnote{\fnsymbol{footnote}}
\setcounter{footnote}{0}

\section{Introduction}

This document describes measurements of normalized differential top-quark-pair (\ttbar) production cross sections done in the dilepton channel using the data sample collected by the CMS experiment~\cite{CMSD}
at a center-of-mass energy of 13 TeV in 2015.
The cross sections are determined as a function of kinematics of the top quarks, \ttbar~system, and as
a function of the number of jets in the event. The measurements, documented in detail in~\cite{CMS13}, are important for performing precision tests of perturbative QCD, and are expected to be sensitive to new phenomena beyond the standard model (SM).

\section{Event selection and kinematic reconstruction}

The \ttbar~events are simulated with the Powheg v2 Monte Carlo (MC) generator interfaced to Pythia 8 for parton showering and hadronization, and using the NNPDF3.0 parton density functions set.
Only prompt-leptonic decays into \ee, \mumu~or \emu~final states are considered for the definition of the \ttbar~signal. 

\begin{figure}[htb]
\centering{
\includegraphics[height=1.88in]{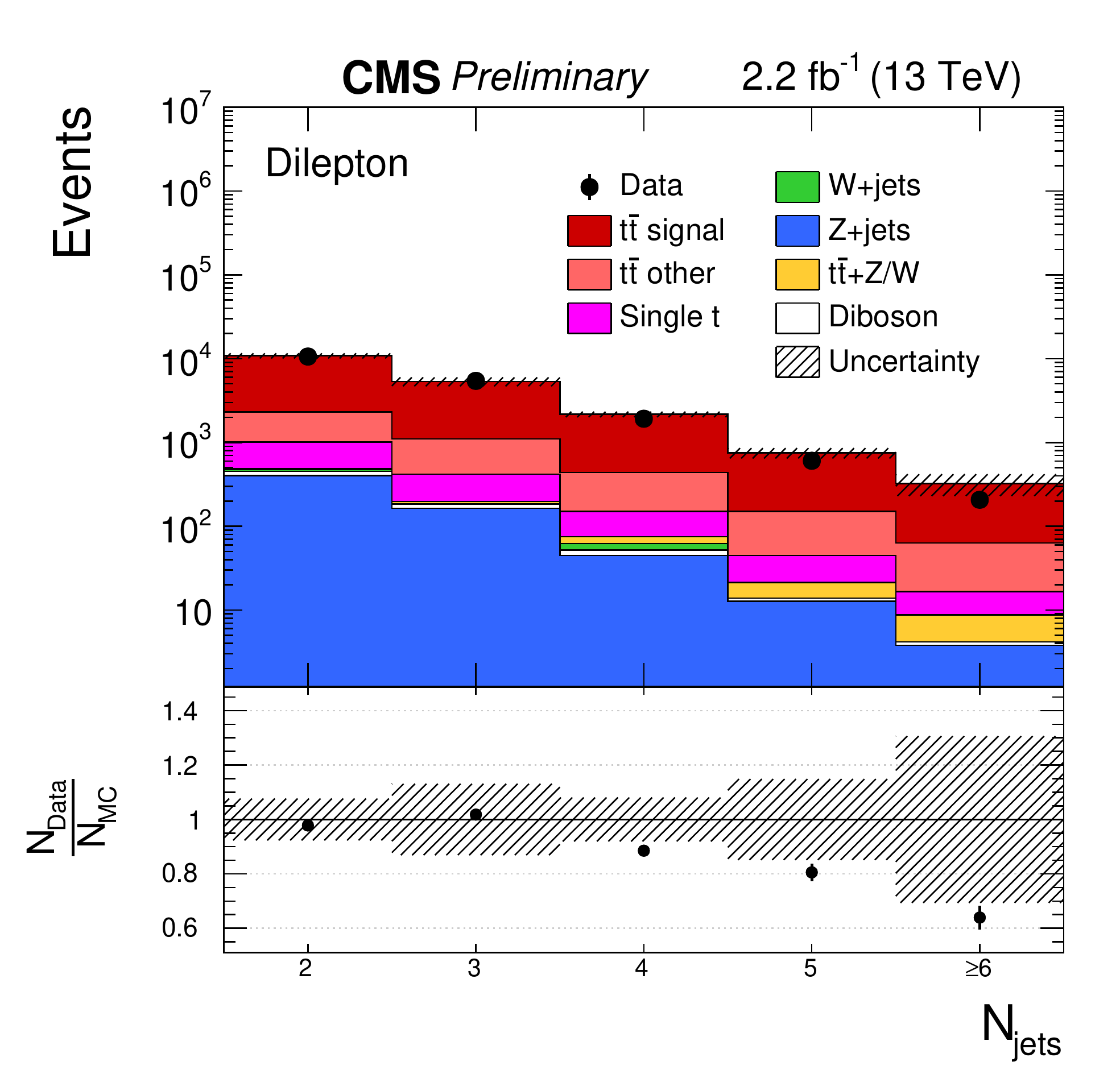}
\includegraphics[height=1.88in]{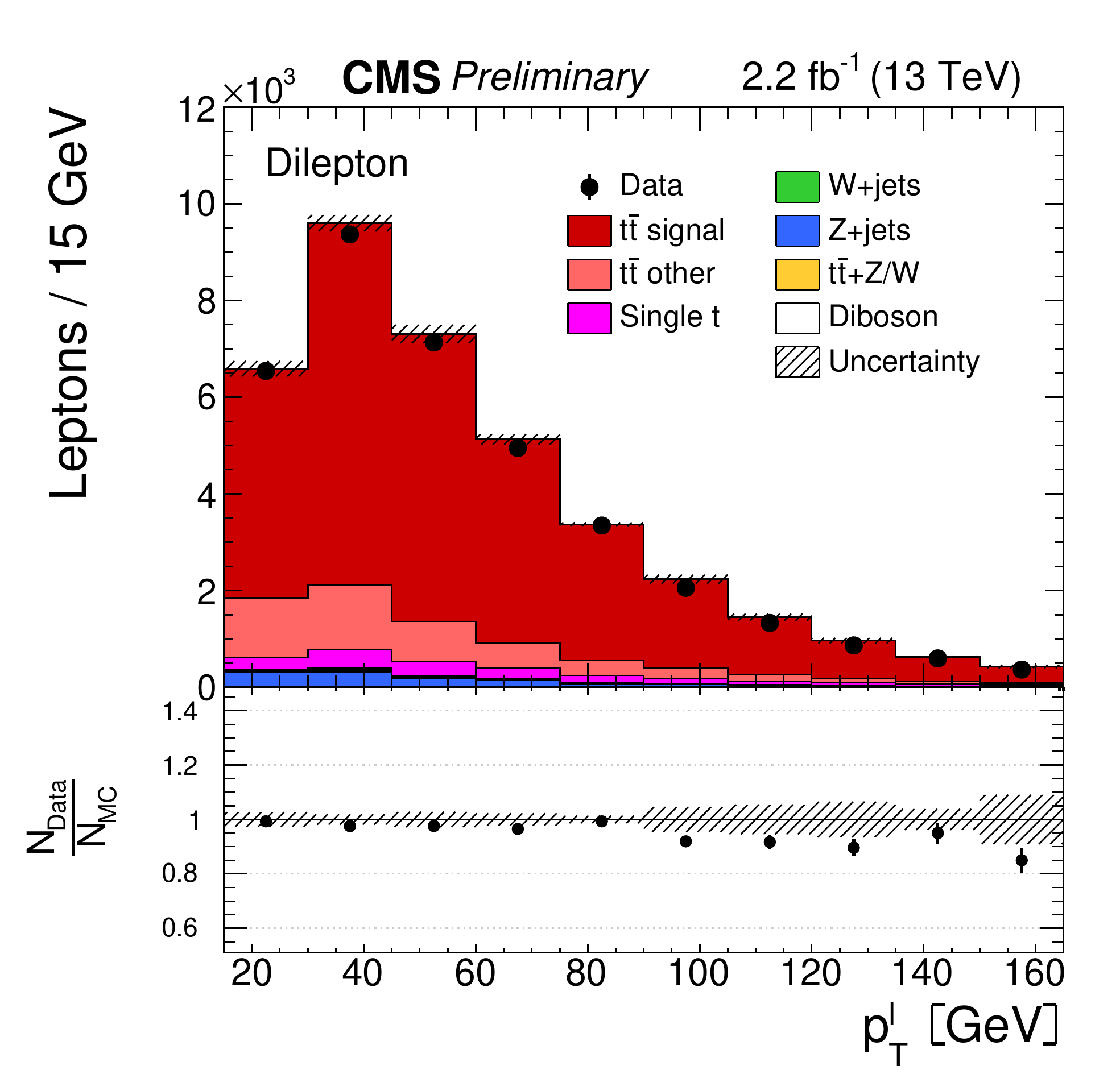}
\includegraphics[height=1.88in]{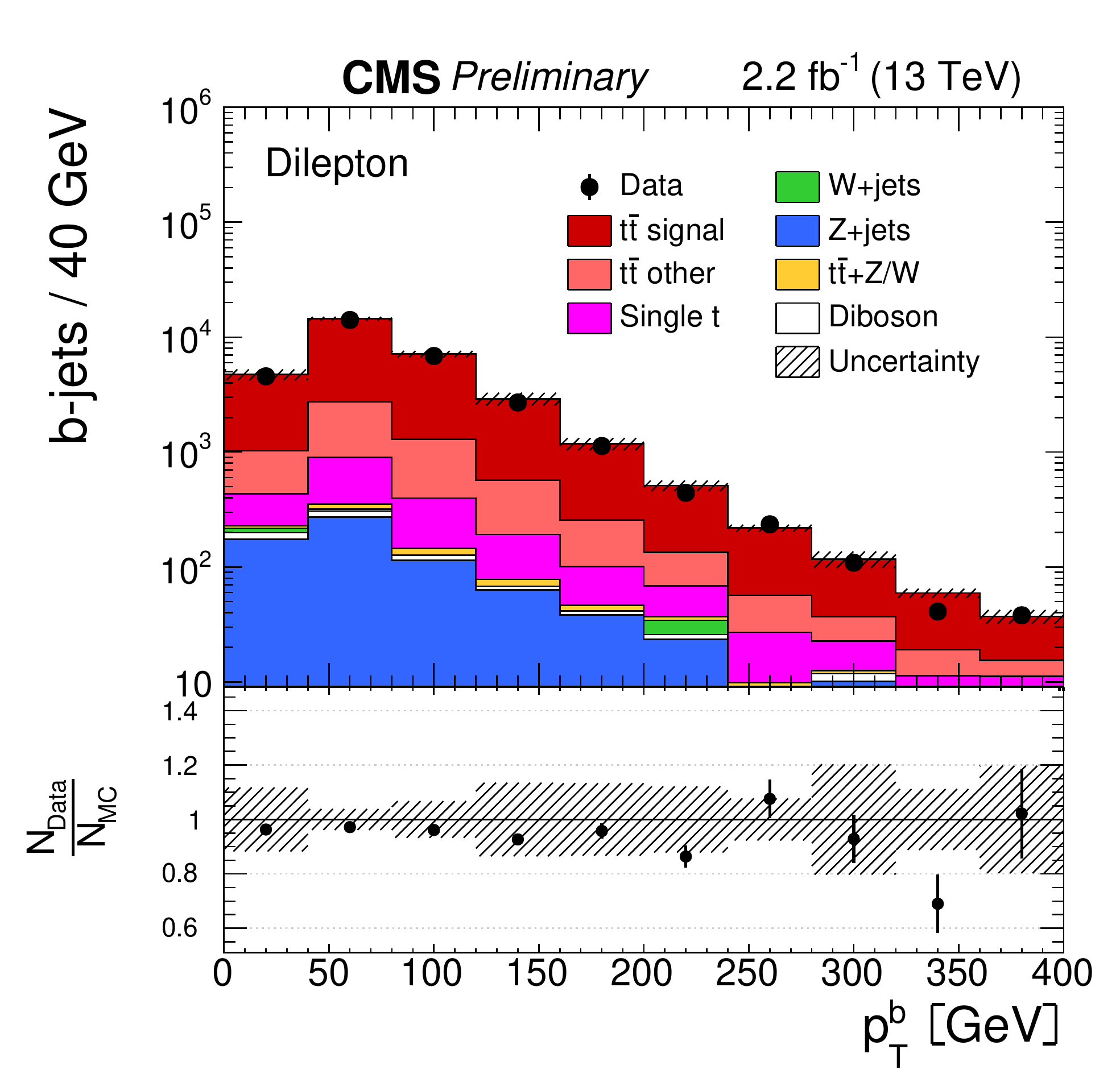}
}
\caption{Spectra of the jet multiplicity~(left), transverse momentum ($p_T$) of the leptons (middle) or b jets (right) after event selection.}
\label{fig:cp_sel}
\end{figure}

Following the well-known topology of a \ttbar~decay in the dilepton channel, the event selection requires exactly two isolated leptons with opposite charge, and at least two energetic jets.
Additional criteria on the invariant mass ($m$) of the dilepton system, b jet identification, and missing transverse energy \met~are applied in order to reduce background contamination. The following background processes
are considered: Z+jets, single top quark (in $tW$-channel), W+jets, diboson (WW, WZ, and ZZ), \ttbar$+$Z$/$W~productions and fakes from other than dilepton \ttbar~decays.

\begin{figure}[htb]
\centering{
\includegraphics[height=1.88in]{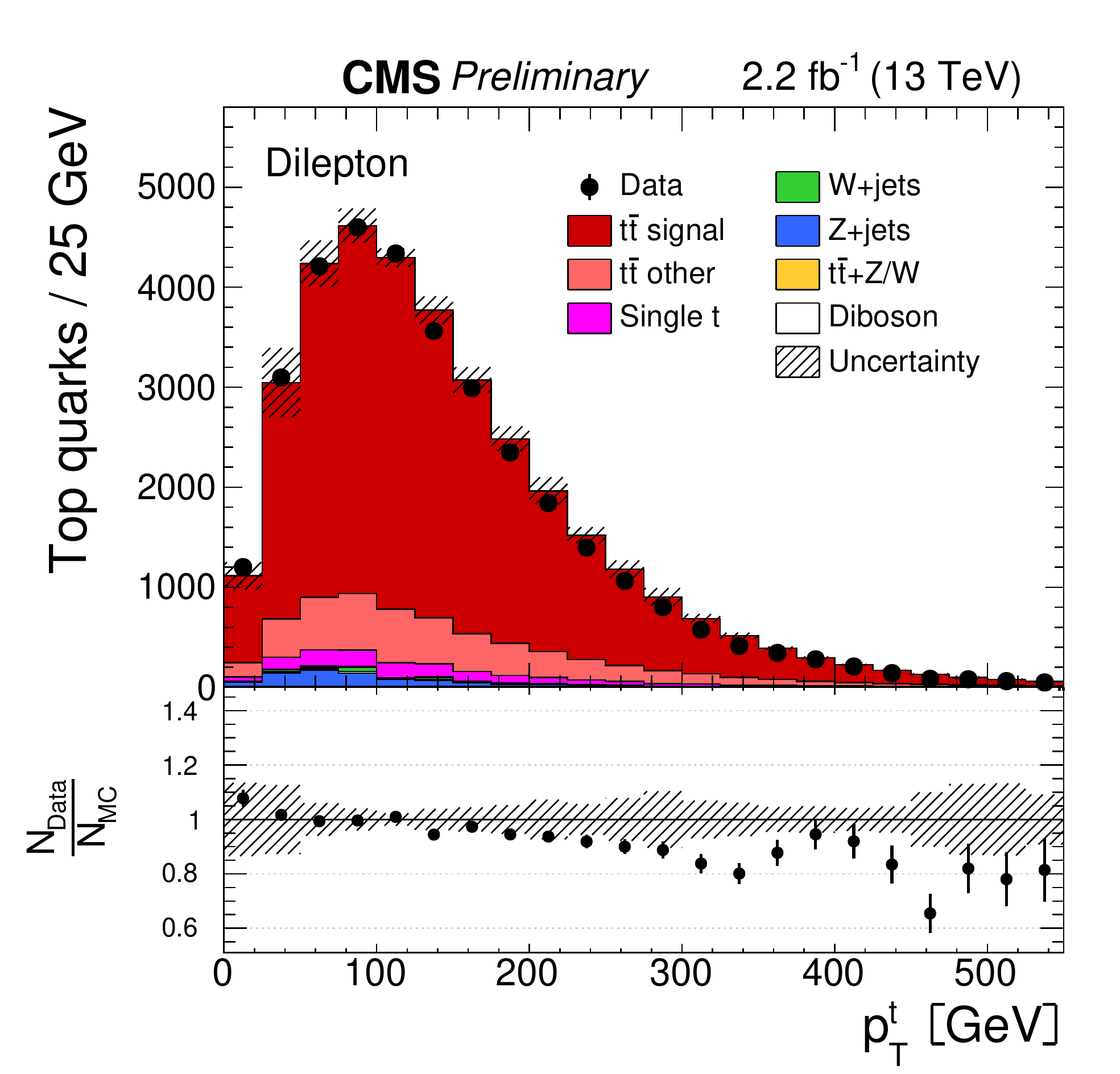}
\includegraphics[height=1.88in]{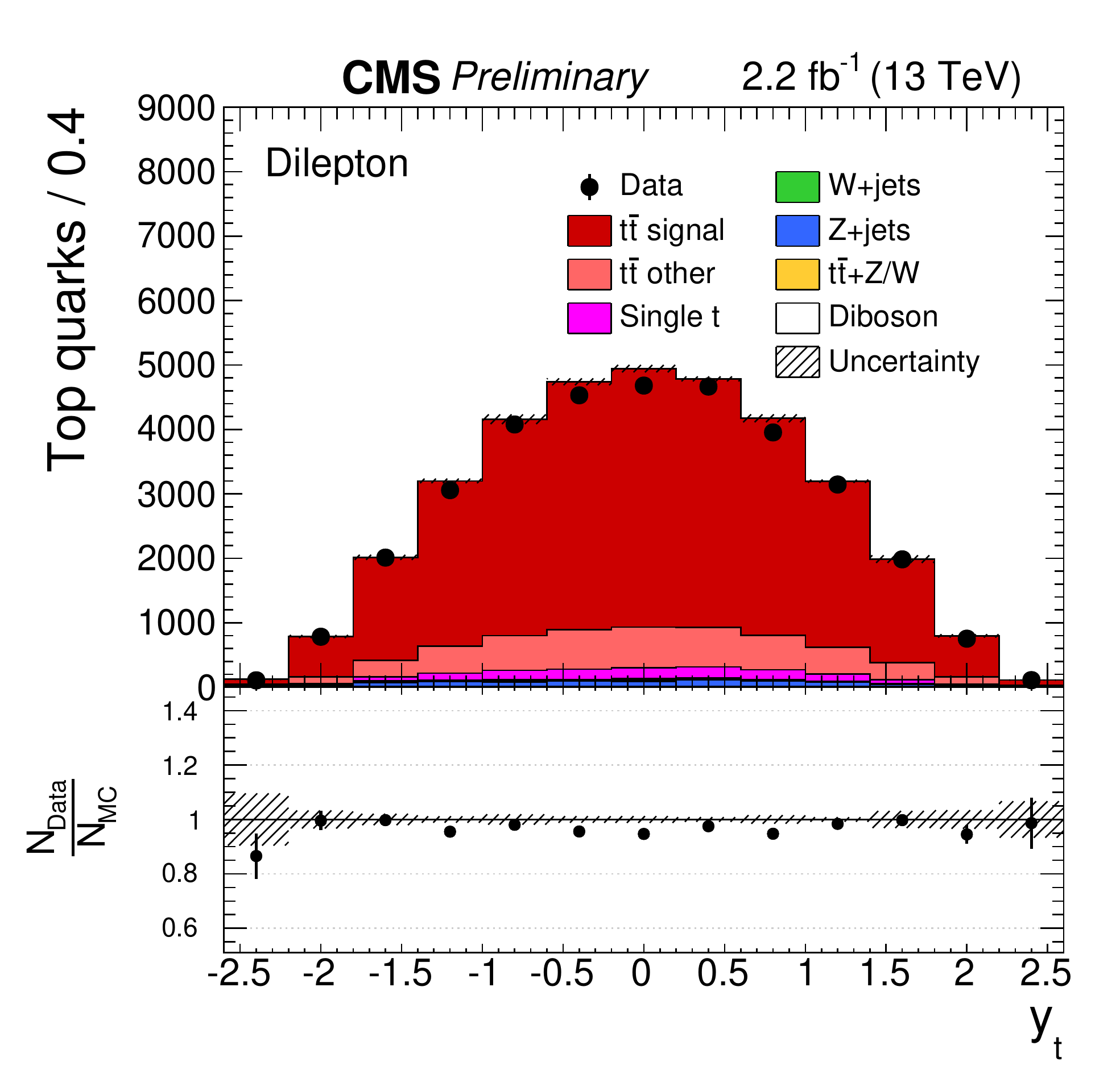}
\includegraphics[height=1.88in]{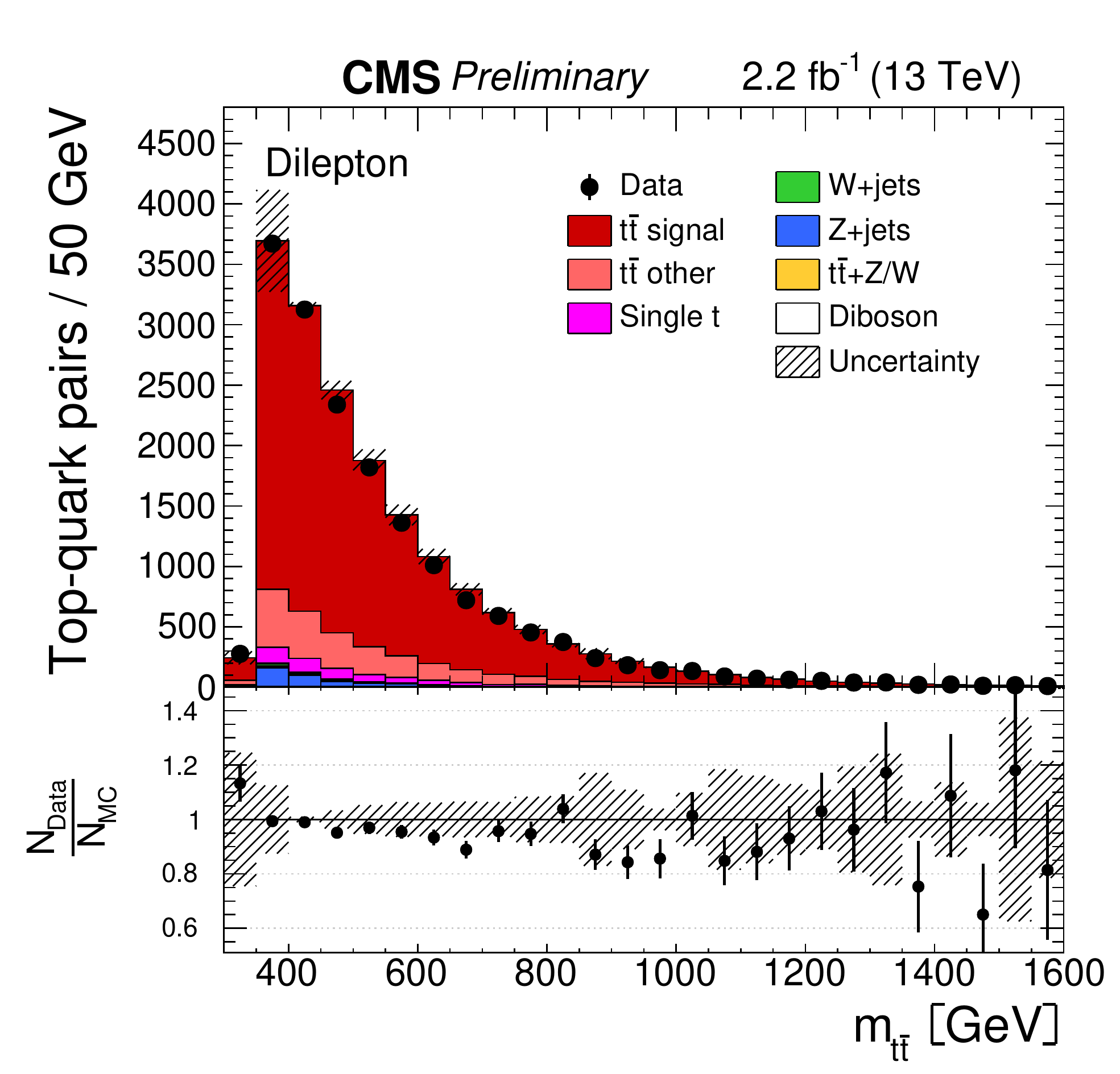}
}
\caption{Spectra of the \toppt~(left), rapidity ($y$) of top quarks~(middle), and \ttm~(right) after kinematic reconstruction.}
\label{fig:cp_top}
\end{figure}

The kinematic properties of the top quarks and \ttbar~system are determined algebraically by solving a system of equations with
respect to the neutrino momenta (six unknowns: three components per $\nu$ or $\bar{\nu}$) using the selected two jets, two leptons, and the \met. The following constraints
are applied: \met$ = p_T^{\nu} + p_T^{\bar{\nu}}$, $m_{W^\pm}$ = 80.4 GeV, and $m_{t,\bar{t}}$ = 172.5 GeV.
Control distributions after full event selection and kinematic reconstruction are shown in Figures~\ref{fig:cp_sel} and~\ref{fig:cp_top}, where
the hatched area indicates shape uncertainties on the \ttbar~signal and backgrounds.

\begin{figure}[htb]
\centering{
\includegraphics[height=2.4in]{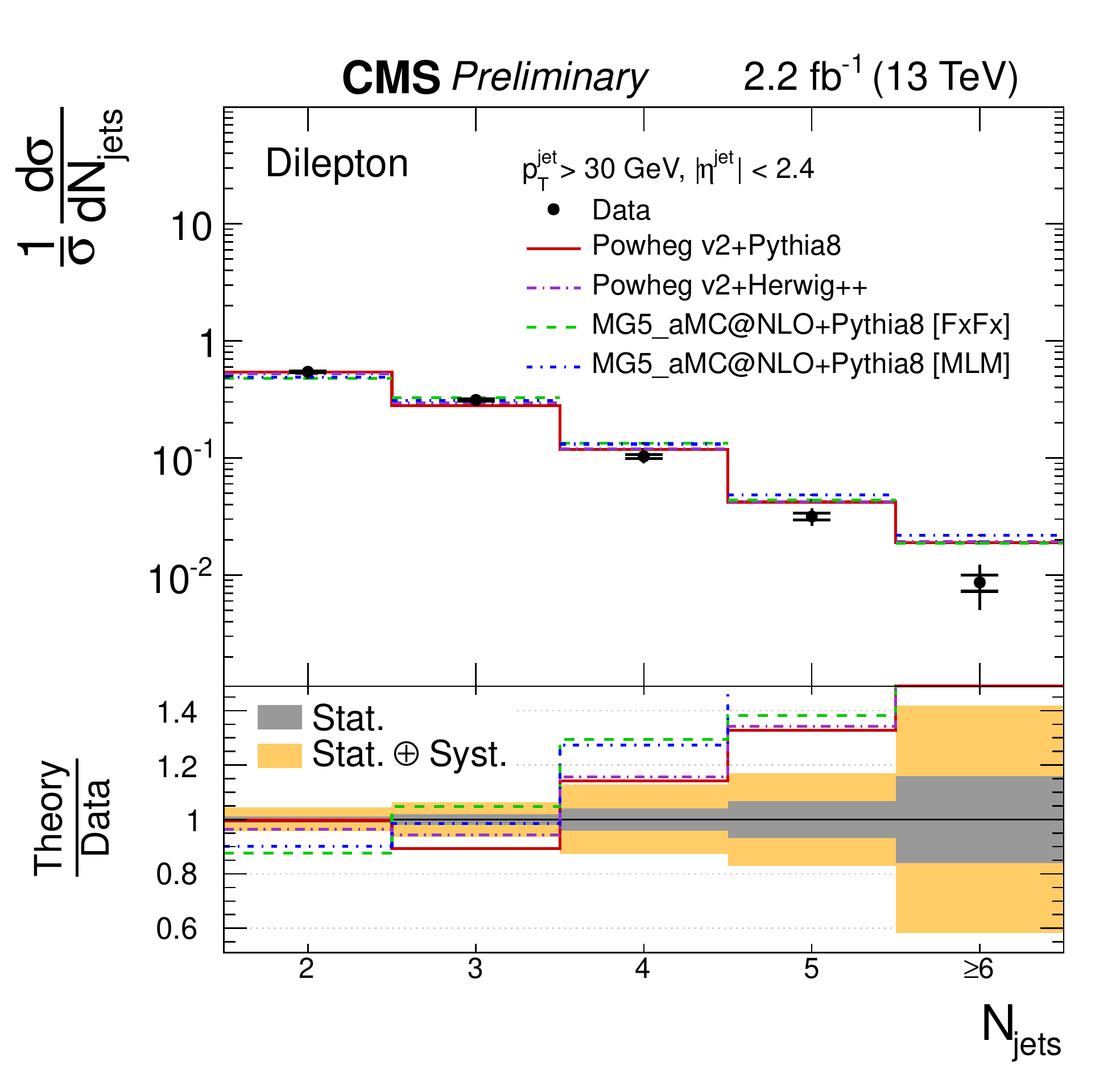}

\includegraphics[height=2.4in]{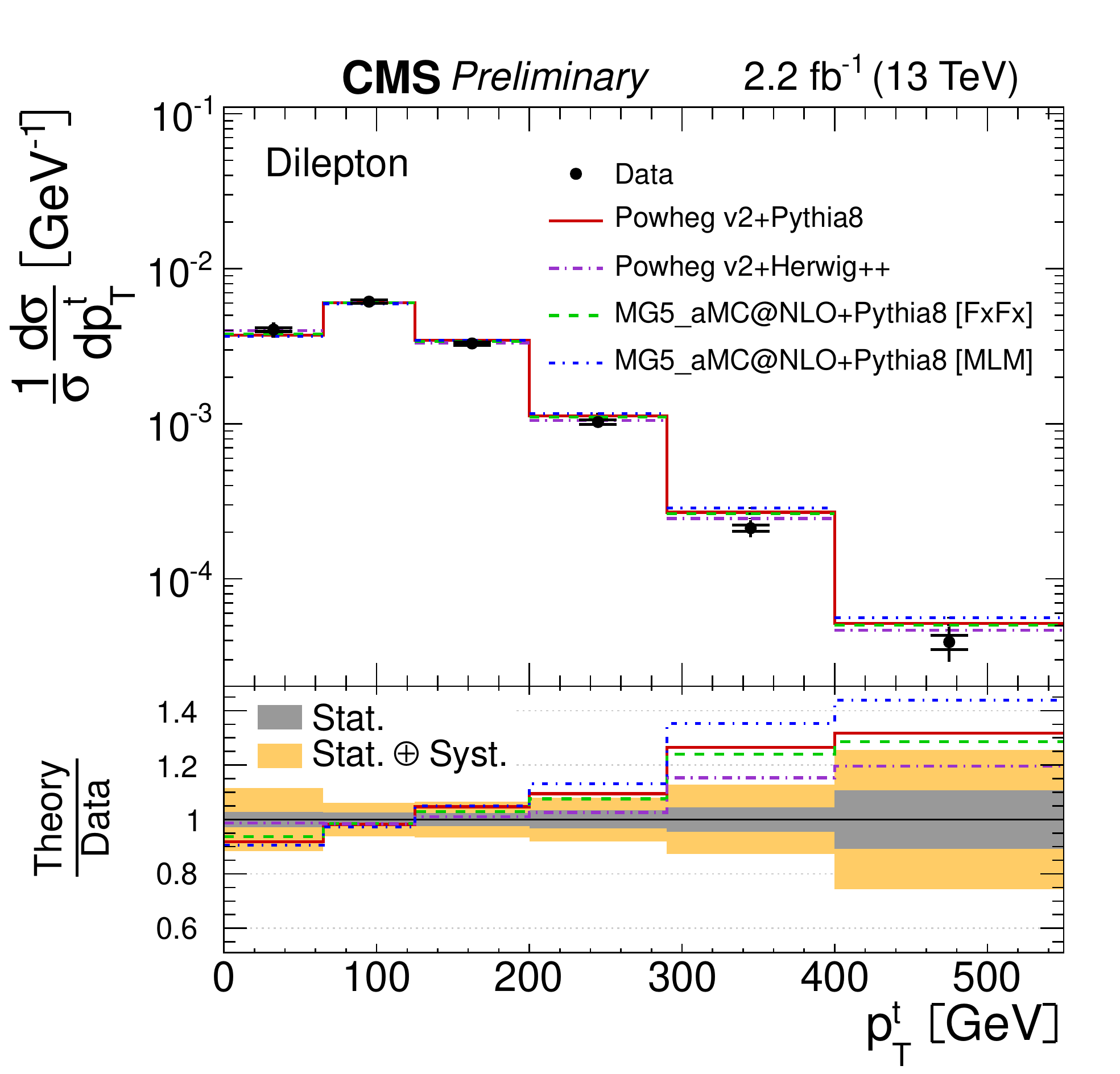}
\includegraphics[height=2.4in]{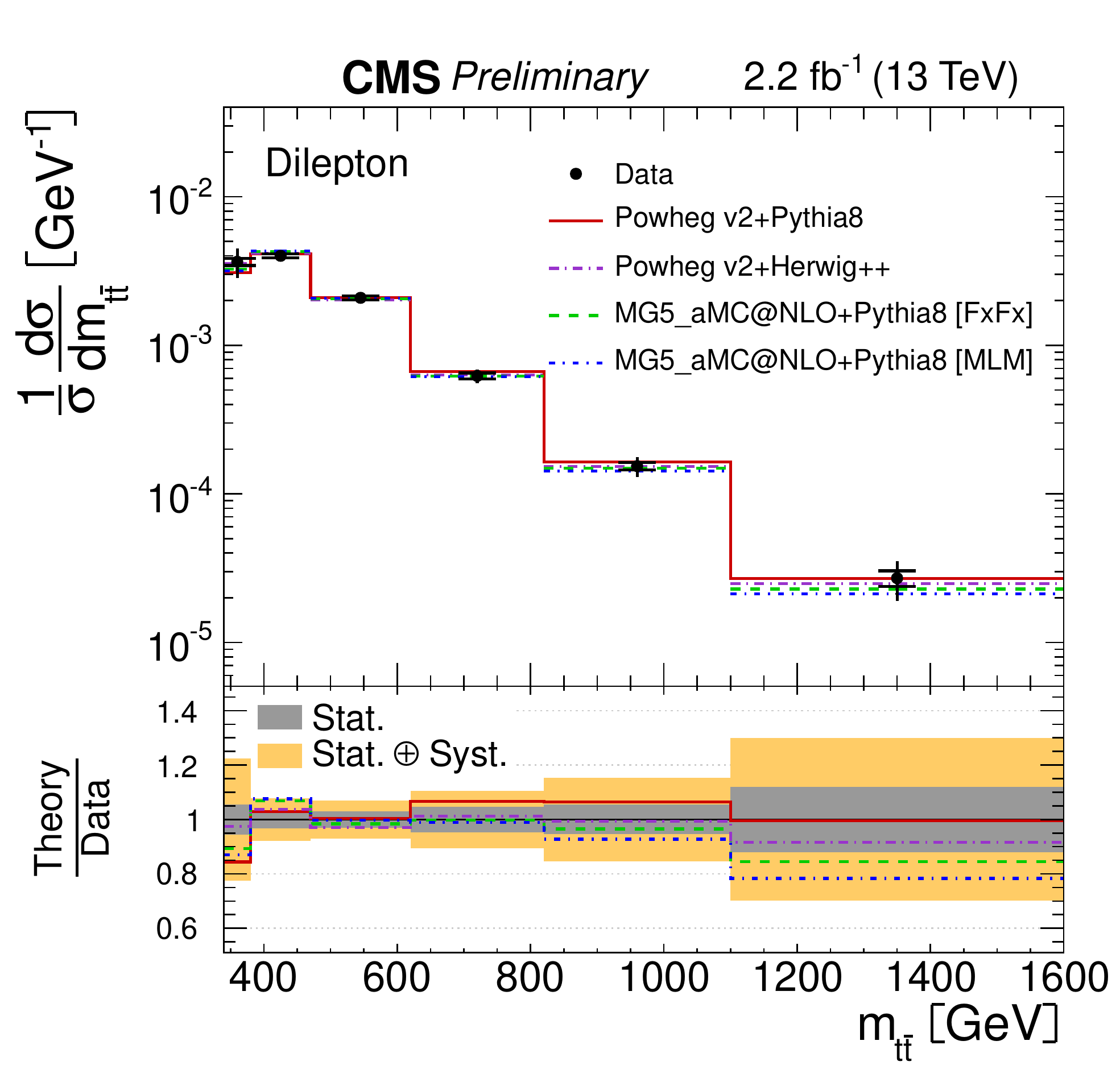}

\includegraphics[height=2.4in]{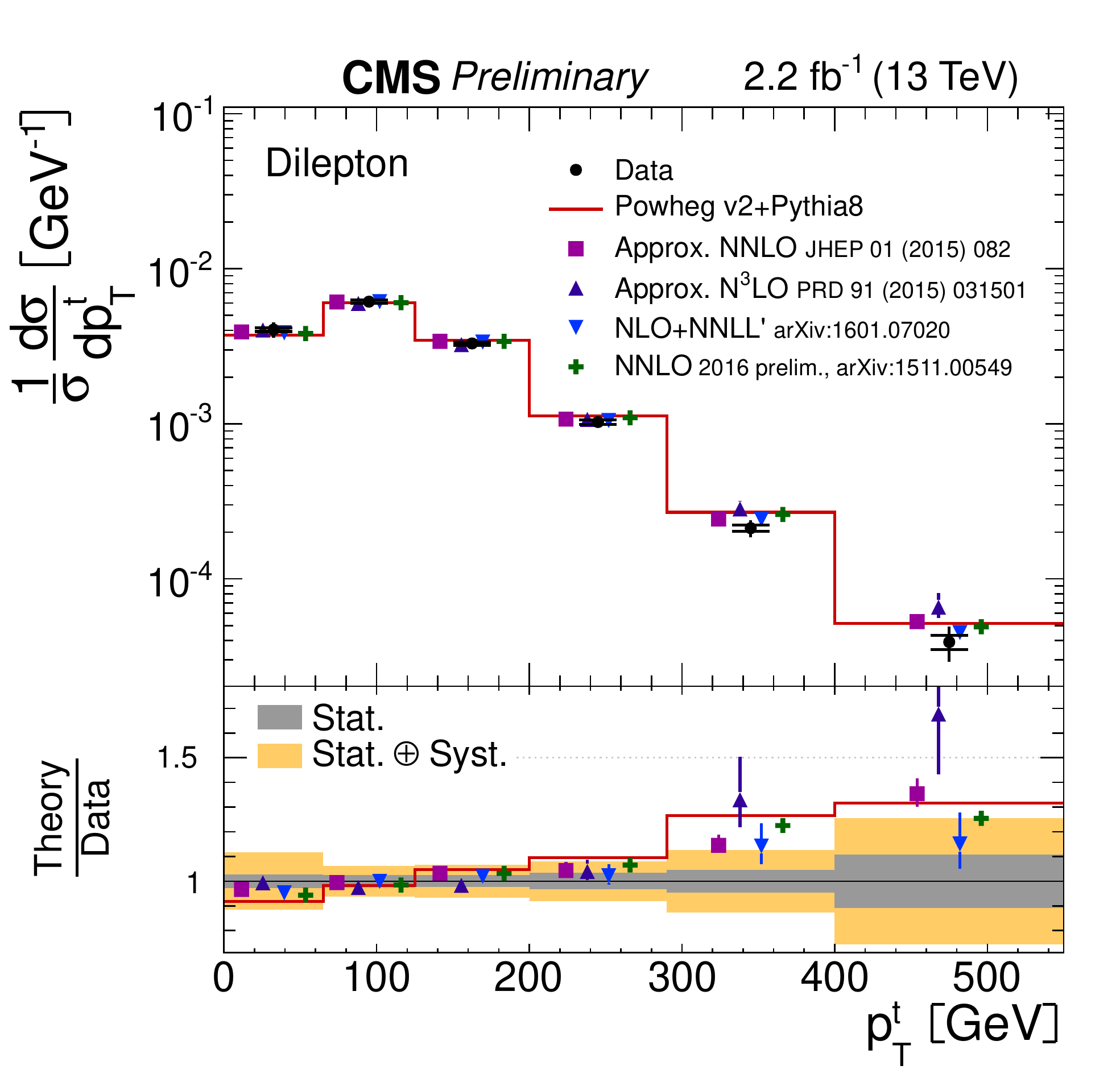}
\includegraphics[height=2.4in]{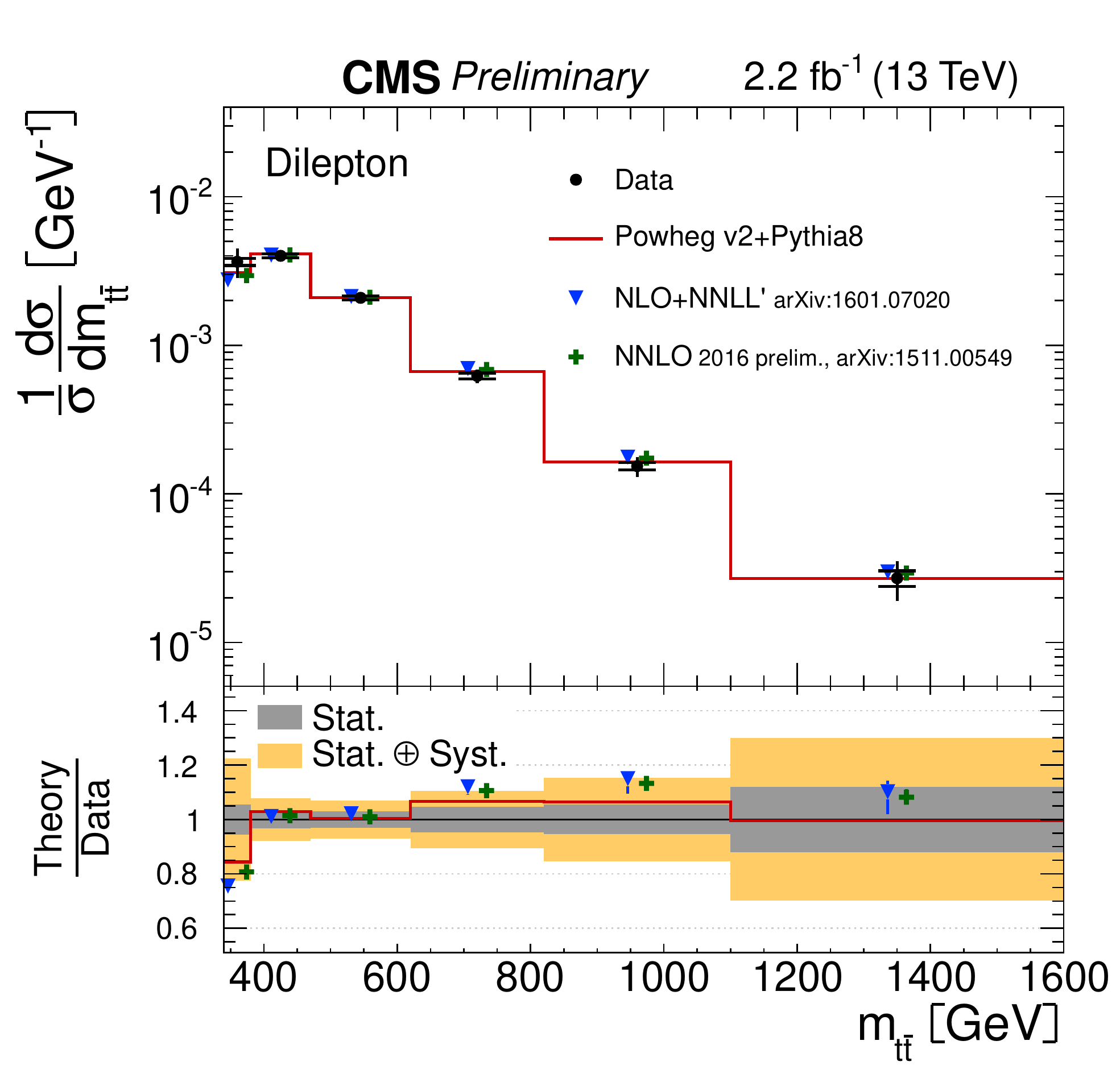}
}
\caption{Normalized differential \ttbar~production cross sections
as a function of the $N_{jets}$ (upper row), \toppt~and \ttm~(middle: compared to MC; bottom: compared to fixed
order calculations). The inner (outer) error bands indicate the statistical (combined statistical
and systematic) uncertainty.}
\label{fig:diff_mc}
\end{figure}

\section{Results}

For a given variable X, the normalized differential cross section is determined as {\large $\frac{1}{\sigma}\cdot\frac{d\sigma_i}{dX_i} = \frac{1}{\sigma}\cdot\frac{x_i}{\Delta^X_i}$},
where $x_i$~respresents the number of signal events observed in data after background subtraction. This quantity is also corrected for the detector efficiencies, acceptances, and migrations across bin boundaries
of the measurement using regularized Singular Value Decomposition unfolding method. The $\Delta^X_i$ is a bin width in units of $X$ and $\sigma$ denotes a measured total
cross section in the visible phase space.

Normalized differential \ttbar~production cross sections measured in bins of the \toppt, \topy, \ttpt, \ttm, \tty~are presented in the full phase space at parton level (i.e. before decay and after QCD radiation).
The measurement performed in bins of the jet multiplicity ($N_{jets}$) is presented in the fiducial phase space,
defined following the event selection requirements, and corrected to particle level to allow for tuning of matrix element + parton shower MC predictions.

A selection of the results are shown in Figure~\ref{fig:diff_mc}, where the measurements are confronted with several modern MC predictions.
The parton level results are also compared with different SM QCD predictions beyond-NLO accuracy~\cite{ANNLO},~\cite{ANNNLO},~\cite{NLONNLL},~\cite{NNLO}.
The overall uncertainty of the measurements ranges from 3 to 30$\%$, with largest contributions from \ttbar~modelling related sources.

\section{Conclusions}

Normalized differential \ttbar~production cross sections are measured at 13 TeV in \pp~collisions
using data corresponding to 2.2 fb$^{-1}$~collected by the CMS detector in 2015.
All measurements are performed in the dilepton decay channel (\ee, \mumu, and \emu).
In general, data are found to be in agreement with modern SM QCD predictions for all measured distributions (\toppt,~\topy,~\ttpt,~\ttm,~\tty~ and $N_{jets}$).
However, the jet multiplicity in data is not very well described by the considered MC simulations for high multiplicity values. After this conference,
further studies, motivated by the results hereby described, have been carried with the purpose
of improving the tune of the current MC predictions~\cite{CMSMC}. The top quark $p_T$~spectrum in data is softer than predictions from MC simulations
and is better described by QCD calculations beyond the NLO-accuracy. This effect was already observed in similar measurements
by the CMS collaboration at 8 TeV~\cite{CMS8}.

\end{document}